# The relation between Pearson's correlation coefficient *r* and Salton's cosine measure

Leo Egghe [1,2] and Loet Leydesdorff [3]

1. Universiteit Hasselt (UHasselt), Campus Diepenbeek, Agoralaan, B-3590 Diepenbeek, Belgium;[1] leo.egghe@uhasselt.be
2. Universiteit Antwerpen (UA), IBW, Stadscampus, Venusstraat 35, B-2000 Antwerpen, Belgium.
3. University of Amsterdam, Amsterdam School of Communication Research (ASCoR), Kloveniersburgwal 48, 1012 CX Amsterdam, The Netherlands; loet@leydesdorff.net

**ABSTRACT**

The relation between Pearson's correlation coefficient and Salton's cosine measure is revealed based on the different possible values of the division of the $L^1$-norm and the $L^2$-norm of a vector. These different values yield a sheaf of increasingly straight lines which form together a cloud of points, being the investigated relation. The theoretical results are tested against the author co-citation relations among 24 informetricians for whom two matrices can be constructed, based on co-citations: the asymmetric occurrence matrix and the symmetric co-citation matrix. Both examples completely confirm the theoretical results. The results enable us to specify an algorithm which provides a threshold value for the cosine above which none of the corresponding Pearson correlations would be negative. Using this threshold value can be expected to optimize the visualization of the vector space.

[1] Permanent address

**Keywords:** Pearson, correlation coefficient, Salton, cosine, non-functional relation, threshold

## 1. Introduction

Ahlgren, Jarneving & Rousseau (2003) questioned the use of Pearson's correlation coefficient as a similarity measure in Author Cocitation Analysis (ACA) on the grounds that this measure is sensitive to zeros. Analytically, the addition of zeros to two variables should add to their similarity, but these authors demonstrated with empirical examples that this addition can depress the correlation coefficient between variables. Salton's cosine is suggested as a possible alternative because this similarity measure is insensitive to the addition of zeros (Salton & McGill, 1983). In general, the Pearson coefficient only measures the degree of a linear dependency. One can expect statistical correlation to be different from the one suggested by Pearson coefficients if a relationship is nonlinear (Frandsen, 2004). However, the cosine does not offer a statistics.

In a reaction White (2003) defended the use of the Pearson correlation hitherto in ACA with the pragmatic argument that the differences resulting from the use of different similarity measures can be neglected in research practice. He illustrated this with dendrograms and mappings using Ahlgren, Jarneving & Rousseau's (2003) own data. Leydesdorff & Zaal (1988) had already found marginal differences between results using these two criteria for the similarity. Bensman (2004) contributed a letter to the discussion in which he argued for the use of Pearson's $r$ for more fundamental reasons. Unlike the cosine, Pearson's $r$ is embedded in multivariate statistics, and because of the normalization implied, this measure allows for negative values.

Jones & Furnas (1987) explained the difference between Salton's cosine and Pearson's correlation coefficient in geometrical terms, and compared both measures with a number of other similarity criteria (Jaccard, Dice, etc.). The Pearson correlation normalizes the values of the vectors to their arithmetic mean. In geometrical terms, this means that the origin of the vector space is located in the middle of the set, while the cosine constructs the vector space from an origin where all vectors have a value of zero (Figure 1).

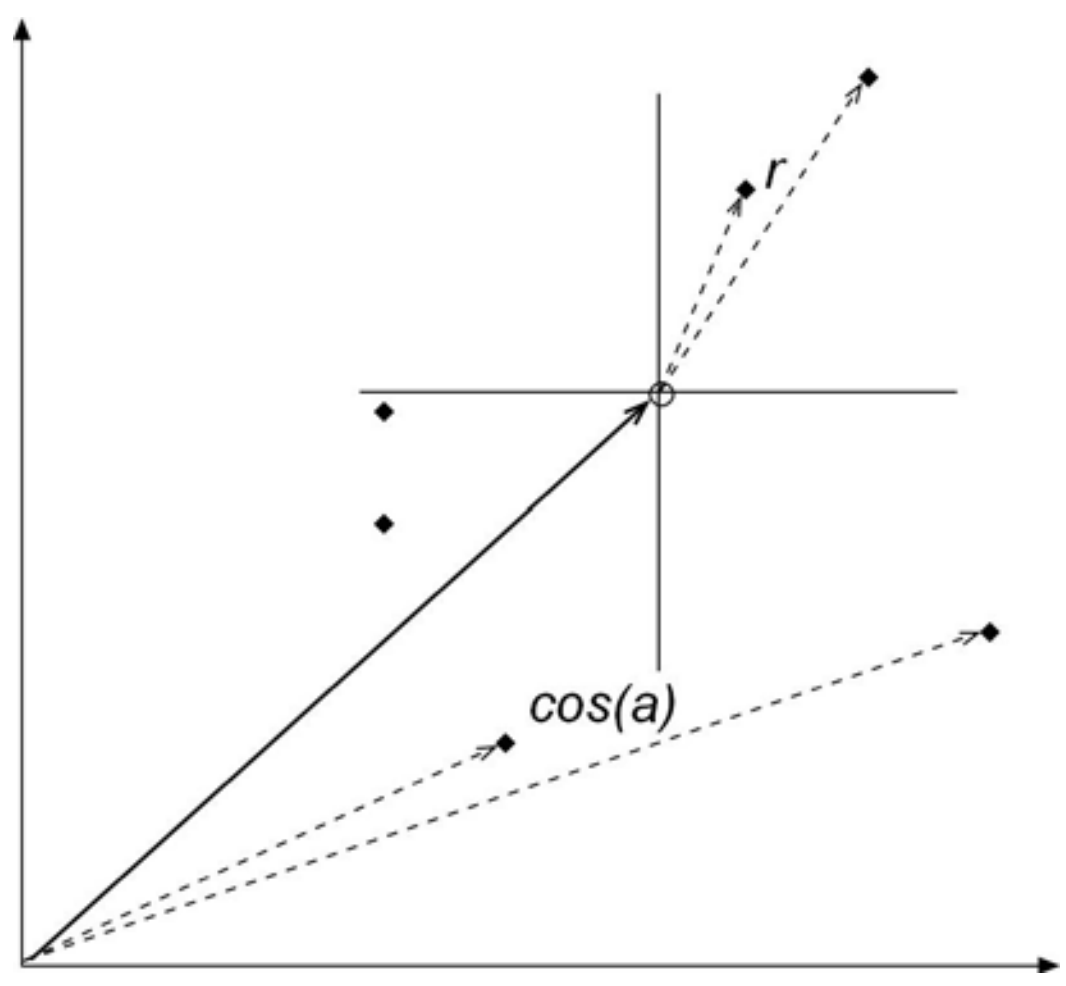

**Figure 1**: The difference between Pearson's *r* and Salton's *cosine* is geometrically equivalent to a translation of the origin to the arithmetic mean values of the vectors.

Consequently, the Pearson correlation can vary from –1 to + 1,[2] while the cosine varies only from zero to one in a single quadrant. In the visualization—using methods based on energy optimization of a system of springs (Kamada & Kawai, 1989) or multidimensional scaling (MDS; see: Kruskal & Wish, 1973; Brandes & Pich, 2007)—this variation in the Pearson correlation is convenient because one can distinguish between positive and negative correlations. Leydesdorff (1986; cf. Leydesdorff & Cozzens, 1993), for example, used this technique to illustrate factor-analytical results of aggregated journal-journal citations matrices with MDS-based journal maps.

Although in many practical cases, the differences between using Pearson's correlation coefficient and Salton's cosine may be negligible, one cannot estimate the significance of this difference in advance. Given the fundamental nature of Ahlgren, Jarneving & Rousseau's (2003, 2004) critique, in our opinion, the cosine is preferable for the analysis and visualization of similarities. Of course, a visualization can be further informed on the basis of multivariate statistics which may very well have to begin with the construction of a Pearson correlation matrix (as in the case of factor analysis). In practice, therefore, one would like to have theoretically informed guidance about choosing the threshold value for the cosine values to be included or not. However, because of the different metrics involved there is no one-to-one correspondence between a cut-off level of *r* = 0 and a value of the cosine similarity.

[2] If one wishes to use only positive values, one can linearly transform the values of the correlation using $(r+1)/2$ (Ahlgren et al., 2003, at p. 552; Leydesdorff and Vaughan, 2006, at p.1617).

Since negative correlations also lead to positive cosine values, the cut-off level is no longer given naturally in the case of the cosine, and, therefore, the choice of a threshold remains somewhat arbitrary (Leydesdorff, 2007a). Yet, variation of the threshold can lead to different visualizations (Leydesdorff & Hellsten, 2006). Using common practice in social network analysis, one could consider using the mean of the lower triangle of the similarity matrix as a threshold for the display (Wasserman & Faust, 1994, at pp. 407f.), but this solution often fails to satisfy the criterion of generating correspondence between, for example, the factor-analytically informed clustering and the clusters visible on the screen.

## 2. Data

Ahlgren, Jarneving & Rousseau (2003 at p. 554) downloaded from the *Web of Science* 430 bibliographic descriptions of articles published in *Scientometrics* and 483 such descriptions published in the *Journal of the American Society for Information Science and Technology (JASIST)* for the period 1996-2000. From the 913 bibliographic references in these articles they composed a co-citation matrix for 12 authors in the field of information retrieval and 12 authors doing bibliometric-scientometric research. They provide both the co-occurrence matrix and the Pearson correlation table in their paper (at p. 555 and 556, respectively).

Leydesdorff & Vaughan (2006) repeated the analysis in order to obtain the original (asymmetrical) data matrix. Using precisely the same searches, these authors found 469 articles in *Scientometrics* and 494 in *JASIST* on 18 November 2004. The somewhat higher numbers are consistent with the practice of Thomson Scientific (ISI) to reallocate papers sometimes at a later date to a previous year. Thus, these differences can be disregarded.

First, we will use the asymmetric occurrence data containing only 0s and 1s: 279 papers contained at least one co-citation to two or more authors on the list of 24 authors under study (Leydesdorff & Vaughan, 2006, p.1620). In this case of an asymmetrical occurrence matrix, an author receives a 1 on a coordinate (representing one of these papers) if he /she is cited in this paper and a score 0 if not. This table is not included here or in Leydesdorff (2008) since it is long (but it can be obtained from the authors upon request).

As a second example, we use the symmetric co-citation data as provided by Leydesdorff (2008, p. 78), Table 1 (as described above). On the basis of this data, Leydesdorff (2008, at p.

78) added the values on the main diagonal to Ahlgren, Jarneving & Rousseau's (2003) Table 7 which provided the author co-citation data (p. 555). The data allows us to compare the various similarity matrices using both the symmetrical co-occurrence data and the asymmetrical occurrence data (Leydesdorff & Vaughan, 2006; Waltman & van Eck, 2007; Leydesdorff, 2007b). This data will be further analyzed after we have established our mathematical model on the relation between Pearson's correlation coefficient *r* and Salton's cosine measure $\text{Cos}$.

## 3. Formalization of the problem

In a recent contribution, Leydesdorff (2008) suggested that in the case of a symmetrical co-occurrence matrix, Small's (1973) proposal to normalize co-citation data using the Jaccard index (Jaccard, 1901; Tanimoto, 1957) has conceptual advantages over the use of the cosine. On the basis of Figure 3 of Leydesdorff (2008, at p. 82), Egghe (2008) was able to show using the same data that all these similarity criteria can functionally be related to one another. The results in Egghe (2008) can be outlined as follows.

Let $\vec{X} = (x_1, x_2, ..., x_n)$ and $\vec{Y} = (y_1, y_2, ..., y_n)$ be two vectors where all the coordinates are positive. The Jaccard index of these two vectors (measuring the "similarity" of these vectors) is defined as

$$J = \frac{\vec{X} \cdot \vec{Y}}{\left\|\vec{X}\right\|_2^2 + \left\|\vec{Y}\right\|_2^2 - \vec{X} \cdot \vec{Y}} \qquad (1)$$

$$J = \frac{\sum_{i=1}^{n} x_i y_i}{\sum_{i=1}^{n} x_i^2 + \sum_{i=1}^{n} y_i^2 - \sum_{i=1}^{n} x_i y_i} \qquad (2)$$

where $\vec{X} \cdot \vec{Y} = \sum_{i=1}^{n} x_i y_i$ is the inproduct of the vectors $\vec{X}$ and $\vec{Y}$ and where $\left\|\vec{X}\right\|_2 = \sqrt{\sum_{i=1}^{n} x_i^2}$ and $\left\|\vec{Y}\right\|_2 = \sqrt{\sum_{i=1}^{n} y_i^2}$ are the Euclidean norms of $\vec{X}$ and $\vec{Y}$ (also called the $L^2$-norms). Salton's cosine measure is defined as

$$\mathrm{Cos} = \frac{\vec{X} \cdot \vec{Y}}{\left\|\vec{X}\right\|_2 \left\|\vec{Y}\right\|_2} \tag{3}$$

$$\mathrm{Cos} = \frac{\sum_{i=1}^{n} x_i y_i}{\sqrt{\sum_{i=1}^{n} x_i^2}\sqrt{\sum_{i=1}^{n} y_i^2}} \tag{4}$$

in the same notation as above. Among other results we could prove that, if $\left\|\vec{X}\right\|_2 = \left\|\vec{Y}\right\|_2$, then

$$J = \frac{\mathrm{Cos}}{2 - \mathrm{Cos}} \tag{5}$$

a simple relation, agreeing completely with the experimental findings.

For Dice's measure E:

$$E = \frac{2\vec{X} \cdot \vec{Y}}{\left\|\vec{X}\right\|_2^2 + \left\|\vec{Y}\right\|_2^2} \tag{6}$$

$$E = \frac{2\sum_{i=1}^{n} x_i y_i}{\sum_{i=1}^{n} x_i^2 + \sum_{i=1}^{n} y_i^2} \tag{7}$$

we could even prove that, if $\left\|\vec{X}\right\|_2 = \left\|\vec{Y}\right\|_2$, we have $E = \mathrm{Cos}$. The same could be shown for several other similarity measures (Egghe, 2008). We refer the reader to some classical monographs which define and apply several of these measures in information science: Boyce, Meadow & Kraft (1995); Tague-Sutcliffe (1995); Grossman & Frieder (1998); Losee (1998); Salton & McGill (1987) and Van Rijsbergen (1979); see also Egghe & Michel (2002, 2003).

Egghe (2008) mentioned the problem of relating Pearson's correlation coefficient with the other measures. The definition of $r$ is:

$$r = \frac{n\sum_{i=1}^{n} x_i y_i - \left(\sum_{i=1}^{n} x_i\right)\left(\sum_{i=1}^{n} y_i\right)}{\sqrt{n\sum_{i=1}^{n} x_i^2 - \left(\sum_{i=1}^{n} x_i\right)^2}\sqrt{n\sum_{i=1}^{n} y_i^2 - \left(\sum_{i=1}^{n} y_i\right)^2}} \tag{8}$$

In this study, we address this remaining question about the relation between Pearson's correlation coefficient and Salton's cosine.

The problem lies in the simultaneous occurrence of the $L^2$-norms of the vectors $\vec{X}=(x_1,...,x_n)$ and $\vec{Y}=(y_1,...,y_n)$ and the $L^1$-norms of these vectors in the definition of the Pearson correlation coefficient. The $L^1$-norms are defined as follows:

$$\left\|\vec{X}\right\|_1=\sum_{i=1}^{n}x_i \tag{9}$$

$$\left\|\vec{Y}\right\|_1=\sum_{i=1}^{n}y_i \tag{10}$$

These $L^1$-norms are the basis for the so-called "city-block metric" (cf. Egghe & Rousseau, 1990). The $L^1$-norms were not occurring in the other measures defined above, and therefore not in Egghe (2008). This makes *r* a special measure in this context. In Ahlgren, Jarneving & Rousseau (2003) argued that *r* lacks some properties that similarity measures should have. Of course, Pearson's *r* remains a very important measure of the degree to which a regression line fits an experimental two-dimensional cloud of points. (See Egghe & Rousseau (2001) for many examples in library and information science.)

Basic for determining the relation between *r* and $\text{Cos}$ will be, evidently, the relation between the $L^1$- and the $L^2$-norms of the vectors $\vec{X}$ and $\vec{Y}$. In the next section we show that every fixed value of $a=\frac{\left\|\vec{X}\right\|_1}{\left\|\vec{X}\right\|_2}$ and of $b=\frac{\left\|\vec{Y}\right\|_1}{\left\|\vec{Y}\right\|_2}$ yields a linear relation between *r* and $\text{Cos}$.

## 4. The mathematical model for the relation between *r* and Cos

Let $\vec{X}=(x_1,x_2,...,x_n)$ and $\vec{Y}=(y_1,y_2,...,y_n)$ the two vectors of length $n$. Denote

$$a=\frac{\left\|\vec{X}\right\|_1}{\left\|\vec{X}\right\|_2} \tag{11}$$

and

$$b=\frac{\left\|\vec{Y}\right\|_1}{\left\|\vec{Y}\right\|_2} \tag{12}$$

(notation as in the previous section). Note that, trivially, $a \geq 1$ and $b \geq 1$. We also have that $a < \sqrt{n}$ and $b < \sqrt{n}$. Indeed, by the inequality of Cauchy-Schwarz (e.g. Hardy, Littlewood & Pólya, 1988) we have

$$\left\|\vec{X}\right\|_1 = \sum_{i=1}^{n} x_i = \sum_{i=1}^{n} 1 \cdot x_i$$

$$\leq \left(\sum_{i=1}^{n} 1\right)^{\frac{1}{2}} \left(\sum_{i=1}^{n} x_i^2\right)^{\frac{1}{2}}$$

$$= \sqrt{n}\left\|\vec{X}\right\|_2$$

Hence

$$a = \frac{\left\|\vec{X}\right\|_1}{\left\|\vec{X}\right\|_2} \leq \sqrt{n}$$

But, if we suppose that $\vec{X}$ is not the constant vector, we have that $a \neq \sqrt{n}$, hence, by the above, $a < \sqrt{n}$. The same argument goes for $\vec{Y}$, yielding $b < \sqrt{n}$. We have the following result.

**<u>Proposition II.1:</u>**

The following relation is generally valid, given (11) and (12) and if $\vec{X}$ nor $\vec{Y}$ are constant vectors

$$r = \frac{n}{\sqrt{n-a^2}\sqrt{n-b^2}}\left(\text{Cos} - \frac{ab}{n}\right) \tag{13}$$

Note that, by the above, the numbers under the roots are positive (and strictly positive neither $\vec{X}$ nor $\vec{Y}$ is constant).

**<u>Proof:</u>**

Define the "Pseudo Cosine" measure PCos

$$\text{PCos} = \frac{\sum_{i=1}^{n} x_i y_i}{\left(\sum_{i=1}^{n} x_i\right)\left(\sum_{i=1}^{n} y_i\right)} \tag{14}$$

One can find earlier definitions in Jones & Furnas (1987). The measure is called "Pseudo Cosine" since, in formula (3) (the real Cosine of the angle between the vectors $\vec{X}$ and $\vec{Y}$, which is well-known), one replaces $\|\vec{X}\|_2$ and $\|\vec{Y}\|_2$ by $\|\vec{X}\|_1$ and $\|\vec{Y}\|_1$, respectively. Hence, as follows from (4) and (14) we have

$$\frac{\mathrm{Cos}}{\mathrm{PCos}} = \frac{\left(\sum_{i=1}^{n} x_i\right)\left(\sum_{i=1}^{n} y_i\right)}{\sqrt{\sum_{i=1}^{n} x_i^2}\sqrt{\sum_{i=1}^{n} y_i^2}}$$

$$\frac{\mathrm{Cos}}{\mathrm{PCos}} = \frac{\|\vec{X}\|_1 \|\vec{Y}\|_1}{\|\vec{X}\|_2 \|\vec{Y}\|_2} = ab, \tag{15}$$

using (11) and (12). Now we have, since neither $\vec{X}$ nor $\vec{Y}$ is constant (avoiding $\frac{0}{0}$ in the next expression).

$$\frac{r}{\mathrm{Cos}} = \frac{n - \frac{\left(\sum_{i=1}^{n} x_i\right)\left(\sum_{i=1}^{n} y_i\right)}{\sum_{i=1}^{n} x_i y_i}}{\sqrt{n - \frac{\left(\sum_{i=1}^{n} x_i\right)^2}{\sum_{i=1}^{n} x_i^2}}\sqrt{n - \frac{\left(\sum_{i=1}^{n} y_i\right)^2}{\sum_{i=1}^{n} y_i^2}}}$$

$$\frac{r}{\mathrm{Cos}} = \frac{n - \frac{1}{\mathrm{PCos}}}{\sqrt{n-a^2}\sqrt{n-b^2}}$$

by (11), (12) and (14). By (15) we now have

$$\frac{r}{\mathrm{Cos}} = \frac{n - \frac{ab}{\mathrm{Cos}}}{\sqrt{n-a^2}\sqrt{n-b^2}}$$

from which $\mathrm{Cos}$ can be resolved:

$$\mathrm{Cos} = \frac{\sqrt{n-a^2}\sqrt{n-b^2}\,r + ab}{n} \tag{16}$$

Since we want the inverse of (16) we have, from (16), that (13) is correct.

Note that (13) is a linear relation between *r* and $\mathrm{Cos}$, but dependent on the parameters $a$ and $b$ (note that $n$ is constant, being the length of the vectors $\vec{X}$ and $\vec{Y}$).

Note that $\mathrm{Cos} = 0$ if and only if

$$r = -\frac{ab}{\sqrt{n-a^2}\sqrt{n-b^2}} < 0 \tag{17}$$

and that $r = 0$ if and only if

$$\mathrm{Cos} = \frac{ab}{n} > 0 \tag{18}$$

Both formulae vary with variable $a$ and $b$, but (17) is always negative and (18) is always positive. Hence, for varying $a$ and $b$, we have obtained a sheaf of increasingly straight lines. Since, in practice, $a$ and $b$ will certainly vary (i.e. the numbers $\frac{\|\vec{X}\|_1}{\|\vec{X}\|_2}$ will not be the same for all vectors) we have proved here that the relation between $r$ and $\mathrm{Cos}$ is not a functional relation (as was the case between all other measures, as discussed in the previous section) but a relation as an increasing cloud of points. Furthermore, one can expect the cloud of points to occupy a range of points, for $\mathrm{Cos} = 0$, below the zero ordinate while, for $r = 0$, the cloud of points will occupy a range of points with positive abscissa values (this is obvious since $\mathrm{Cos} \geq 0$ while all vector coordinates are positive). Note also that (17) (its absolute value) and (18) decrease with $n$, the length of the vector (for fixed $a$ and $b$). This is also the case for the slope of (13), going, for large $n$, to 1, as is readily seen (for fixed $a$ and $b$).

All these findings will be confirmed in the next section where exact numbers will be calculated and compared with the experimental graphs.

## 5. One example and two applications

As noted, we re-use the reconstructed data set of Ahlgren, Jarneving & Rousseau (2003) which was also used in Leydesdorff (2008). This data deals with the co-citation features of 24 informetricians. We distinguish two types of matrices (yielding the different vectors representing the 24 authors).

First, we use the binary asymmetric occurrence matrix: a matrix of size 279 x 24 as described in section 2. Then, we use the symmetric co-citation matrix of size 24 x 24 where the main diagonal gives the number of papers in which an author is cited – see Table 1 in Leydesdorff (2008, at p. 78). Although these matrices are constructed from the same data set, it will be clear that the corresponding vectors are very different: in the first case all vectors have binary values and length $n = 279$; in the second case the vectors are not binary and have length $n = 24$. So these two examples will also reveal the n-dependence of our model, as described above.

*5.1 The case of the binary asymmetric occurrence matrix*

Here $n = 279$. Hence the model (13) (and its consequences such as (17) and (18)) are known as soon as we have the values $a$ and $b$ as in (11) and (12), i.e., we have to know the values $\frac{\|\vec{X}\|_1}{\|\vec{X}\|_2}$ for every author, represented by $\vec{X}$. Since all vectors are binary we have, for every vector $\vec{X}$:

$$\frac{\|\vec{X}\|_1}{\|\vec{X}\|_2} = \frac{\text{sum of the 1s (ones) in } \vec{X}}{\sqrt{\text{sum of the 1s (ones) in } \vec{X}}}$$

$$\frac{\|\vec{X}\|_1}{\|\vec{X}\|_2} = \sqrt{\text{sum of the 1s (ones) in } \vec{X}} \tag{19}$$

We have the data as in Table 1. They are nothing other than the square roots of the main diagonal elements in Table 1 in Leydesdorff (2008).

**Table 1.** $\frac{\|\vec{X}\|_1}{\|\vec{X}\|_2}$ **for the 24 authors**

| Author | $\frac{\|\vec{X}\|_1}{\|\vec{X}\|_2}$ ($a$ or $b$ in (13)) |
|---|---|
| Braun | $\sqrt{50}$ |
| Schubert | $\sqrt{60}$ |
| Glänzel | $\sqrt{53}$ |
| Moed | $\sqrt{55}$ |
| Nederhof | $\sqrt{31}$ |
| Narin | $\sqrt{64}$ |
| Tyssen | $\sqrt{22}$ |
| van Raan | $\sqrt{50}$ |
| Leydesdorff | $\sqrt{46}$ |
| Price | $\sqrt{54}$ |
| Callon | $\sqrt{26}$ |
| Cronin | $\sqrt{24}$ |
| Cooper | $\sqrt{30}$ |
| Van Rijsbergen | $\sqrt{30}$ |
| Croft | $\sqrt{18}$ |
| Robertson | $\sqrt{36}$ |
| Blair | $\sqrt{18}$ |
| Harman | $\sqrt{31}$ |
| Belkin | $\sqrt{36}$ |
| Spink | $\sqrt{21}$ |
| Fidel | $\sqrt{23}$ |
| Marchionini | $\sqrt{24}$ |
| Kuhltau | $\sqrt{26}$ |
| Dervin | $\sqrt{20}$ |

For (13) we do not need the a- and $b$-values of all authors: to see the range of the $r$-values, given a $\text{Cos}$-value we only calculate (13) for the two smallest and largest values for $a$ and $b$.

1. Smallest values: $a=\sqrt{18}$, $b=\sqrt{20}$

   yielding $ab=\sqrt{360}=18.973666$
2. Largest values: $a=\sqrt{64}$, $b=\sqrt{60}$

   yielding $ab=\sqrt{3,840}=61.967734$

This is a rather rough argument: not all a- and b-values occur at every fixed $\text{Cos}$-value so that better approximations are possible, but for the sake of simplicity we will use the larger margins above: if we can approximate the experimental graphical relation between $r$ and $\text{Cos}$ in a satisfactory way, the model is approved.

Using (13), (17) or (18) we obtain, in each case, the range in which we expect the practical $(\text{Cos}, r)$ points to occur. For $\text{Cos}=0$ we have *r* between $-0.0729762$ and $-0.2869153$ (by (17)). For $r=0$ we have by (18), $\text{Cos}$ between $0.068006$ and $0.2221066$. Further, by (13), for $\text{Cos}=0.1$ we have *r* between $0.0343323$ and $-0.15$. For $\text{Cos}=0.2$ we have *r* between $0.1416408$ and $-0.028424$. For $\text{Cos}=0.3$ we have *r* between $0.2489421$ and $0.1001529$. Finally for $\text{Cos}=0.4$ we have *r* between $0.3562577$ and $0.2287298$ and for $\text{Cos}=0.5$ we have *r* between $0.4635662$ and $0.3573067$. We do not go further due to the scarcity of the data points.

The experimental $(\mathrm{Cos}, \mathrm{r})$ cloud of points and the limiting ranges of the model are shown together in Fig. 2, so that the comparison is easy.

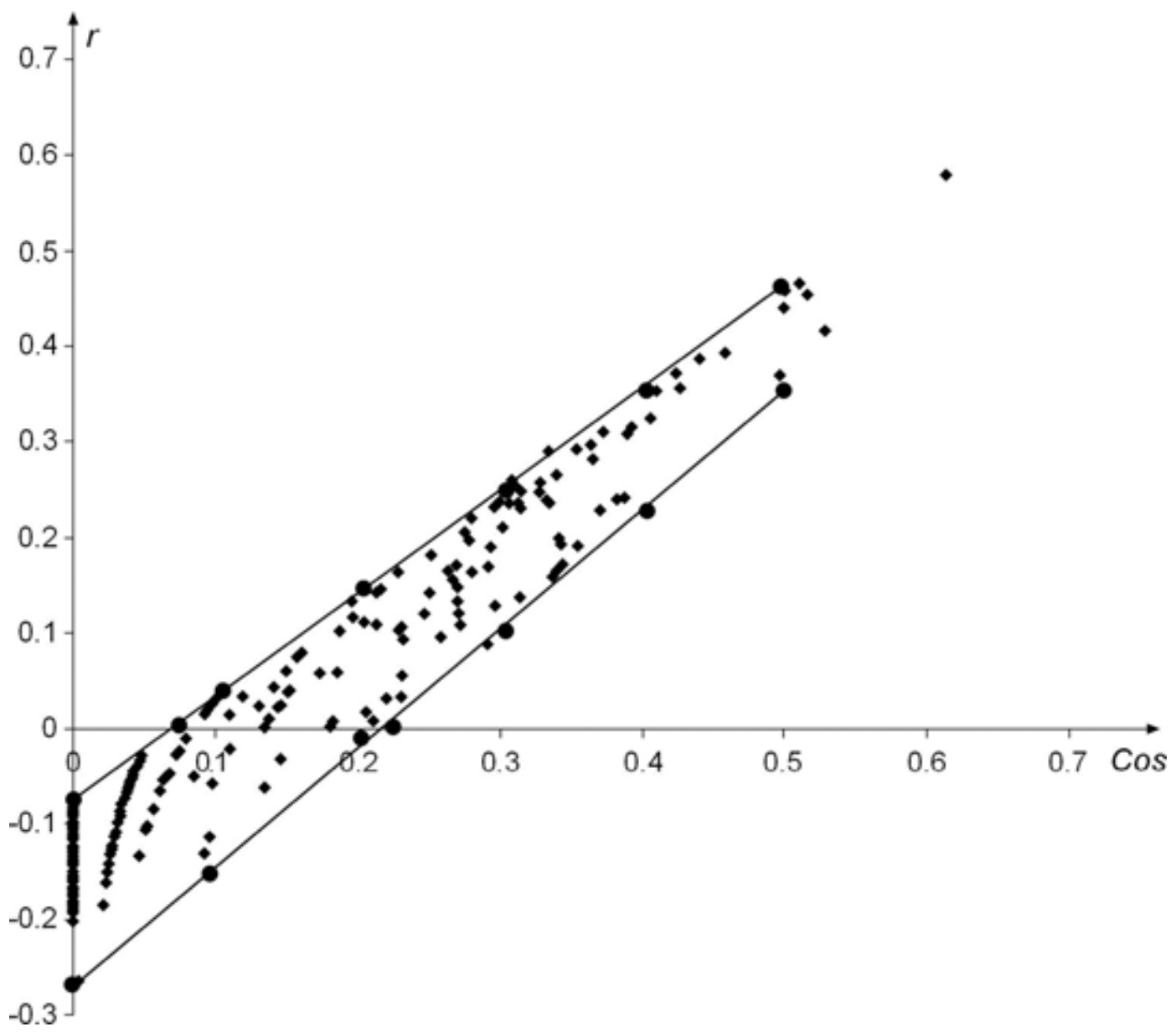

**Figure 2:** Data points $(\mathrm{Cos}, \mathrm{r})$ for the binary asymmetric occurrence matrix and ranges of the model.

For reasons of visualization we have connected the calculated ranges. Figure 2 speaks for itself. The indicated straight lines are the upper and lower lines of the sheaf of straight lines composing the cloud of points. The higher the straight line, the smaller its slope. The *r*-range (thickness) of the cloud decreases as $\mathrm{Cos}$ increases. We also see that the negative *r*-values, e.g. at $\mathrm{Cos} = 0$, are explained, although the lowest fitted point on $\mathrm{Cos} = 0$ is a bit too low due to the fact that we use the total $a, b$ range while, on $\mathrm{Cos} = 0$, not all a- and b-values occur.

We can say that the model (13) explains the obtained $(\mathrm{Cos}, \mathrm{r})$ cloud of points. We will now do the same for the other matrix. We will then be able to compare both clouds of points and both models.

## *5.2 The case of the symmetric co-citation matrix*

Here $n = 24$. Based on Table 1 in Leydesdorff (2008), we have the values of $\frac{\|\vec{X}\|_1}{\|\vec{X}\|_2}$. For example, for "Braun" in the first column of this table, $\|\vec{X}\|_1 = \sum_{i=1}^{n} x_i = 168$ and $\|\vec{X}\|_2 = \sqrt{\sum_{i=1}^{n} x_i^2} = \sqrt{4,504} = 67.1118469$. In this case, $\frac{\|\vec{X}\|_1}{\|\vec{X}\|_2} = 168 / 67.1118469 = 2.5032838$.

The values of $\frac{\|\vec{X}\|_1}{\|\vec{X}\|_2}$ for all 24 authors, represented by their respective vector $\vec{X}$, are provided in Table 2.

**Table 2: $\frac{\|\vec{X}\|_1}{\|\vec{X}\|_2}$ for the 24 authors**

| Author | $\frac{\|\vec{X}\|_1}{\|\vec{X}\|_2}$ ($a$ or $b$ in (13)) |
|---|---|
| Braun | 2.5032838 |
| Schubert | 2.4795703 |
| Glänzel | 2.729457 |
| Moed | 2.7337391 |
| Nederhof | 2.8221626 |
| Narin | 2.8986697 |
| Tyssen | 3.0789273 |
| van Raan | 2.4077981 |
| Leydesdorff | 2.8747094 |
| Price | 2.7635278 |
| Callon | 2.8295923 |
| Cronin | 2.556743 |
| Cooper | 2.3184046 |
| Van Rijsbergen | 2.4469432 |
| Croft | 3.0858543 |
| Robertson | 2.920658 |
| Blair | 2.517544 |
| Harman | 2.5919129 |
| Belkin | 2.8555919 |
| Spink | 3.0331502 |
| Fidel | 2.6927563 |
| Marchionini | 2.4845716 |
| Kuhltau | 2.4693658 |
| Dervin | 2.5086617 |

As in the previous example, we only use the two smallest and largest values for $a$ and $b$.

1. Smallest values: $a = 2.3184046$, $b = 2.4077981$
   yielding $ab = 5.5822502$
2. Largest values: $a = 3.0858543$, $b = 3.0789273$
   yielding $ab = 9.501121$

As in the first example, the obtained ranges will probably be a bit too large, since not all a- and b-values occur at every $\mathrm{Cos}$-value. We will now investigate the quality of the model in this case.

If $\mathrm{Cos} = 0$ then, by (17) we have that *r* is between $-0.3031765$ and $-0.6553024$. If $r = 0$ we have that $\mathrm{Cos}$ is between $0.2325928$ and $0.39588$, using (18). For $\mathrm{Cos} = 0.1$ we have that *r* is between $-0.1728293$ and $-0.4897716$. For $\mathrm{Cos} = 0.2$, *r* is between $-0.0424834$ and $-0.3242411$. $\mathrm{Cos} = 0.4$ implies that *r* is between $0.2182085$ and $0.0068199$. $\mathrm{Cos} = 0.6$ implies that *r* is between $0.4789003$ and $0.3378808$ and finally, for $\mathrm{Cos} = 0.8$ we have that *r* is between $0.7395922$ and $0.6689418$.

The experimental $(\mathrm{Cos}, \mathrm{r})$ cloud of points and the limiting ranges of the model in this case are shown together in Figure 3.

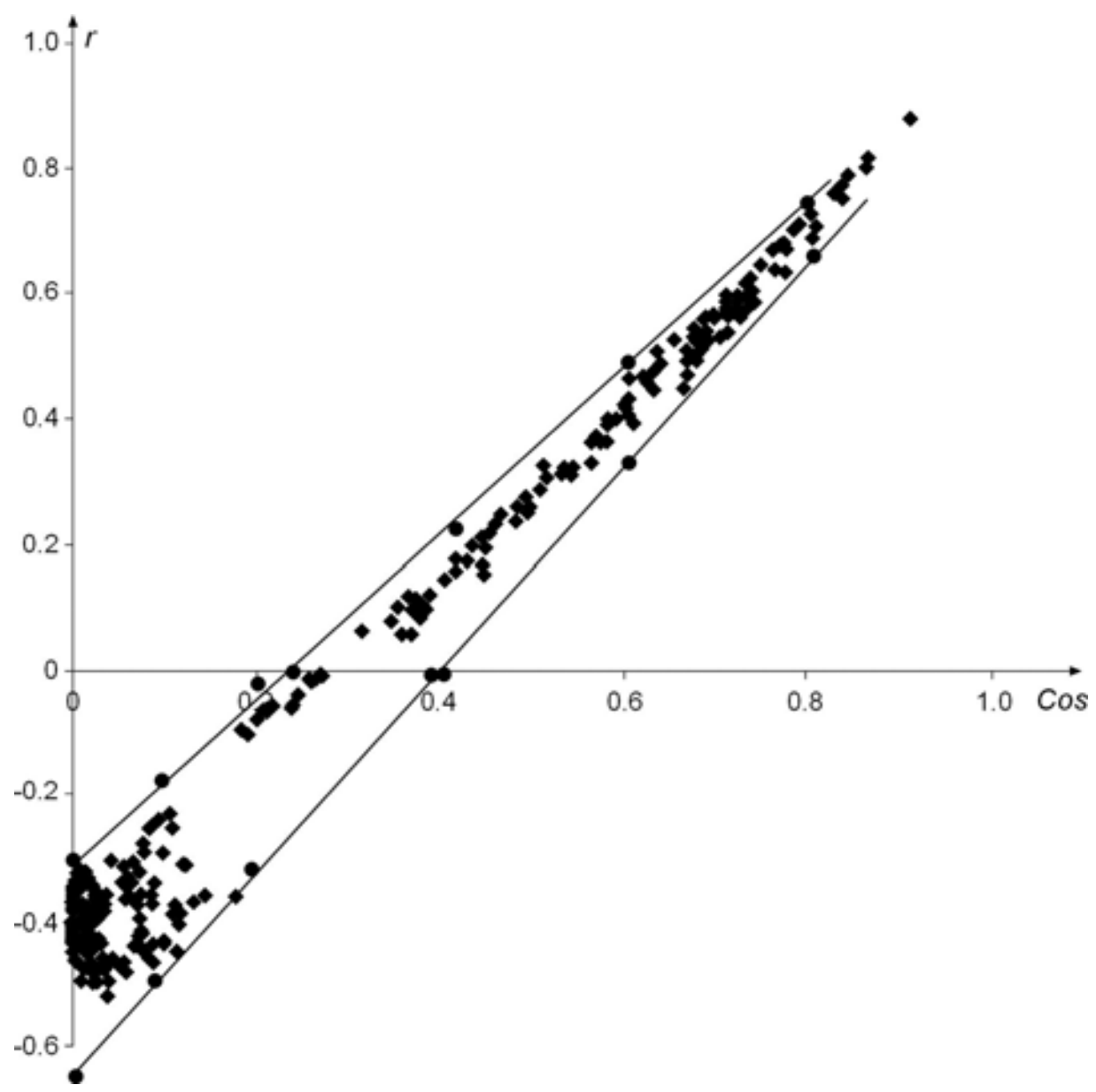

**Figure 3:** Data points $(\mathrm{Cos}, \mathrm{r})$ for the symmetric co-citation matrix and ranges of the model.

The same properties are found here as in the previous case, although the data are completely different. Again the lower and upper straight lines, delimiting the cloud of points, are clear. They also delimit the sheaf of straight lines, given by (13). Again, the higher the straight line, the smaller its slope. The *r*-range (thickness) of the cloud decreases as $\mathrm{Cos}$ increases. This effect is stronger in Fig. 3 than in Fig. 2. We again see that the negative values of *r*, e.g. at $\mathrm{Cos} = 0$, are explained.

We conclude that the model (13) explains the obtained $(\mathrm{Cos}, \mathrm{r})$ cloud of points.

## 6. The effects of the predicted threshold values on the visualization

Figure 4 provides a visualization using the asymmetrical matrix (n = 279) and the Pearson correlation for the normalization.[3] Negative values for the Pearson correlation are indicated with dashed edges.

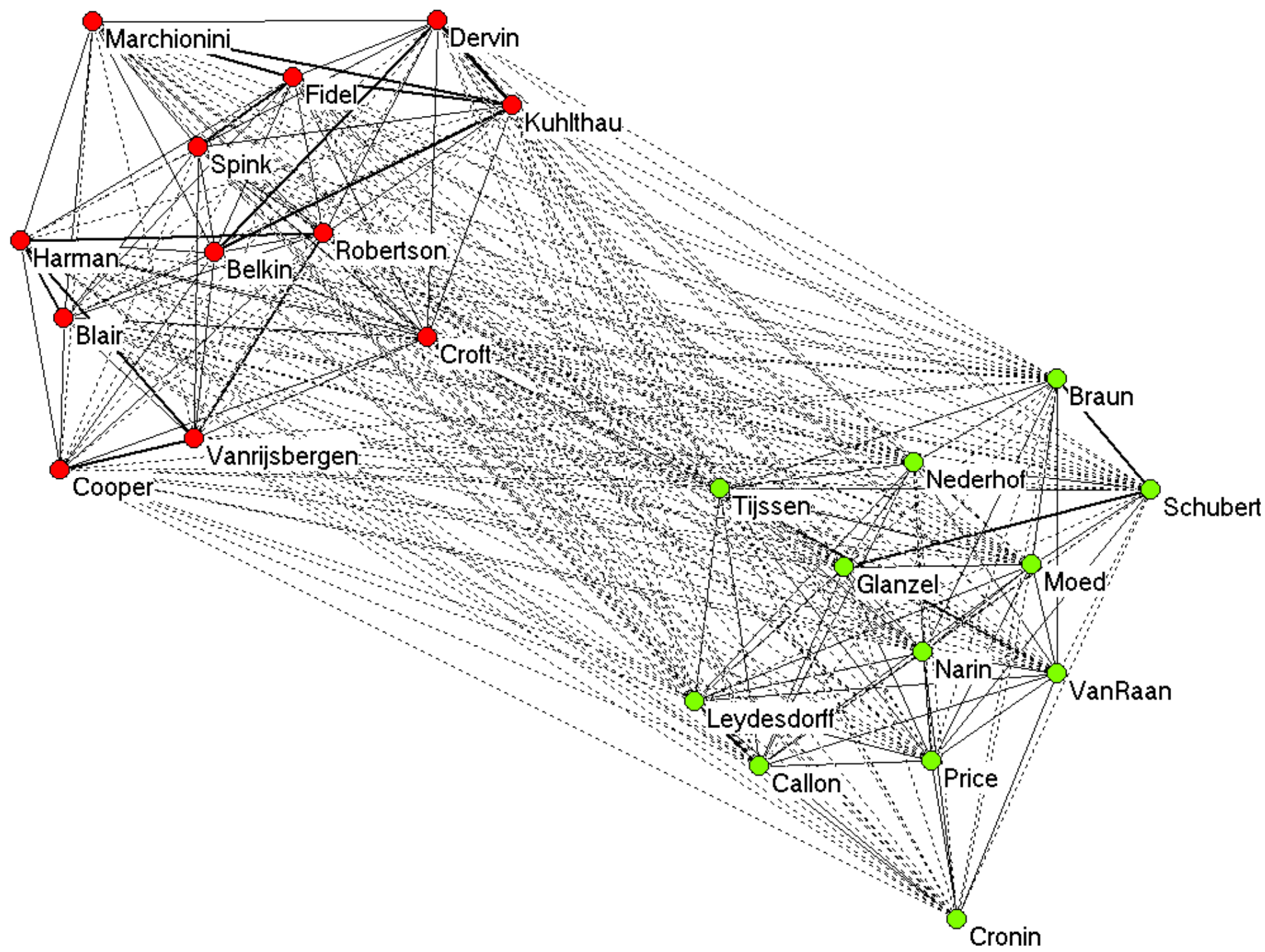

[3] We use the asymmetrical occurrence matrix for this demonstration because it can be debated whether co-occurrence data should be normalized for the visualization (Leydesdorff & Vaughan, 2008; Waltman & Van Eck, 2008; Leydesdorff, 2007b).

**Figure 4**: Pearson correlation among citation patterns of 24 authors in the information sciences in 279 citing documents.

Only positive correlations are indicated within each of the two groups with the single exception of a correlation (*r* = 0.031) between the citation patterns of "Croft" and "Tijssen." This *r* = 0.031 accords with cosine = 0.101. In section 5.1, it was shown that given this matrix (n = 279), *r* = 0 ranges for the cosine between 0.068 and 0.222. Figure 2 (above) showed that several points are within this range. However, there are also negative values for *r within* each of the two main groups. For example, "Cronin" has positive correlations with only five of the twelve authors in the group on the lower right side: "Narin" (*r* = 0.11), "Van Raan" (*r* = 0.06), "Leydesdorff" (*r* = 0.21), "Callon" (*r* = 0.08), and "Price" (*r* = 0.14). All other correlations of "Cronin" are negative.

If we use the lower limit for the threshold value of the cosine (0.068), we obtain Figure 5.

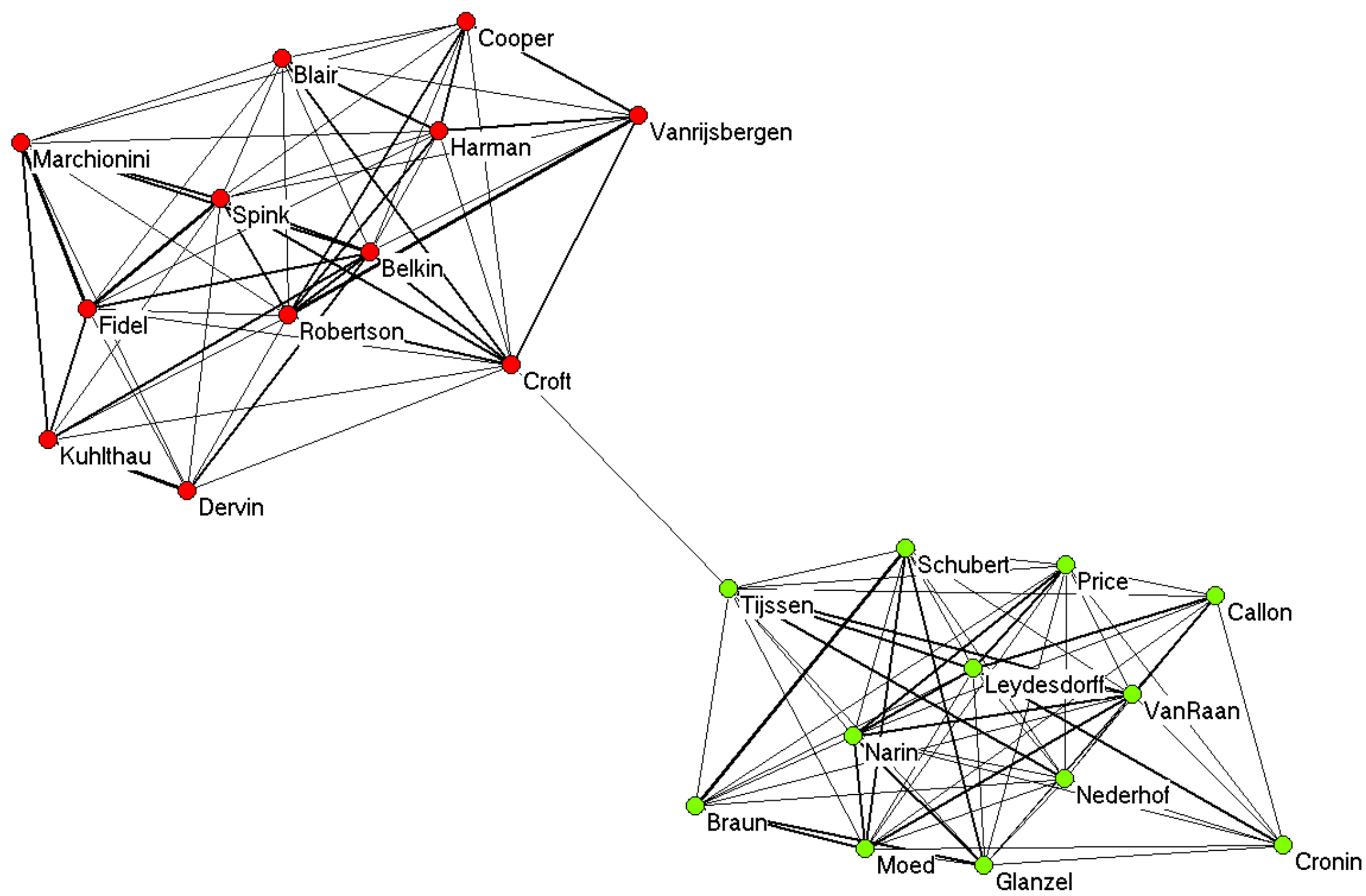

**Figure 5**: Visualization of the same matrix based on cosine > 0.068.

The two groups are now separated, but connected by the one positive correlation between "Tijssen" and "Croft". This is fortunate because this correlation is above the threshold value. In addition to relations to the five author names correlated positively to "Cronin", however, "Cronin" is in this representation erroneously connected to "Moed" ($r = -0.02$), "Nederhof" ($r = -0.03$), and "Glanzel" ($r = -0.05$).

Figure 6 provides the visualization using the upper limit of the threshold value (0.222).

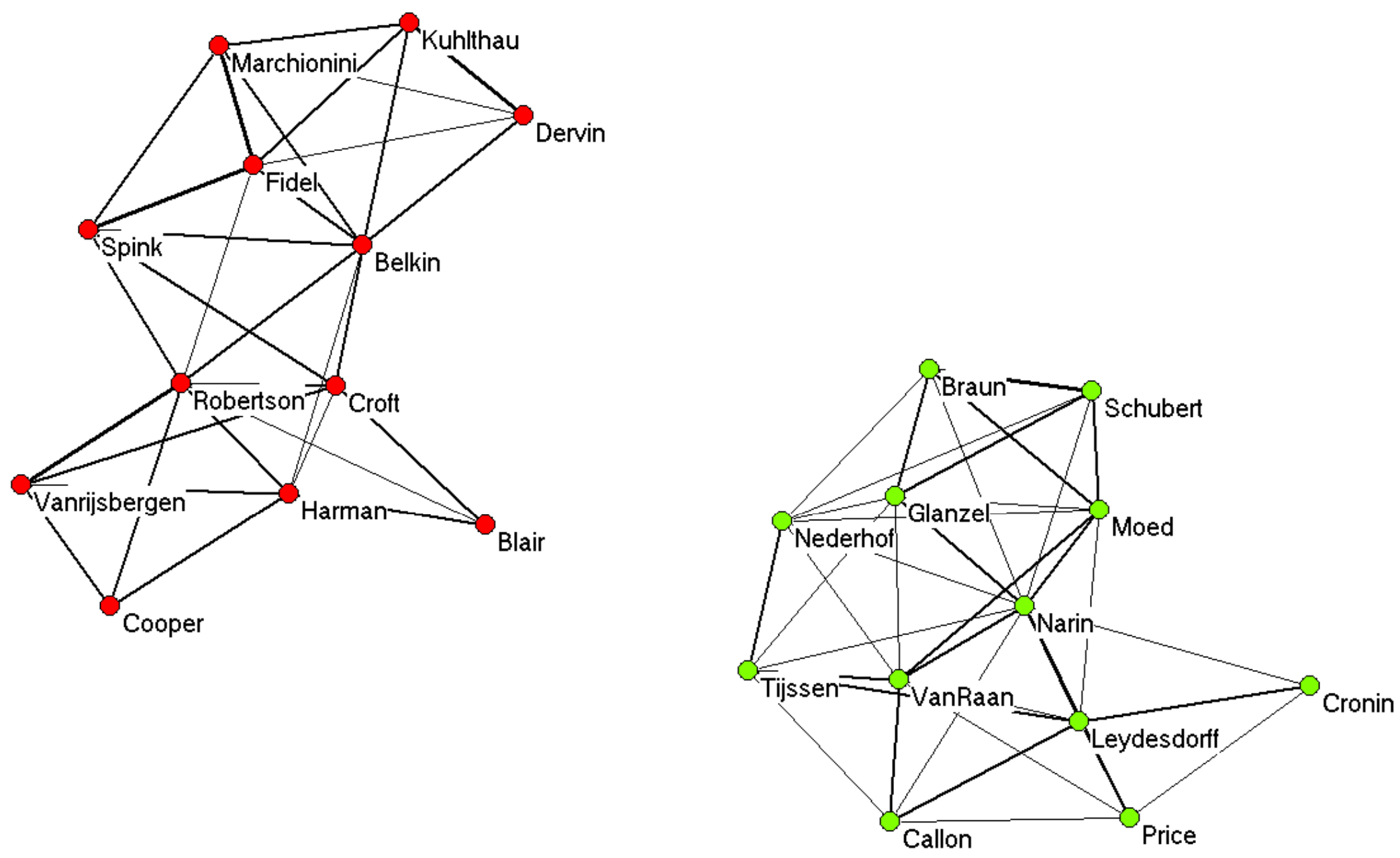

**Figure 6**: Visualization of the same matrix based on cosine > 0.222.

In this visualization, the two groups are no longer connected, and thus the correlation between "Croft" and "Tijssen" ($r = 0.31$) is not appreciated. Similarly, the correlation of "Cronin" with two other authors at a level of $r < 0.1$ ("Van Raan" and "Callon") is no longer visualized. However, all correlations at the level of $r > 0.1$ are made visible. (Since these two graphs are independent, the optimization using Kamada & Kawai's (1989) algorithm was repeated.) The graphs are additionally informative about the internal structures of these communities of authors. Using this upper limit of the threshold value, in summary, prevents the drawing of edges which correspond with negative correlations, but is conservative. This is a property which one would like in most representations.

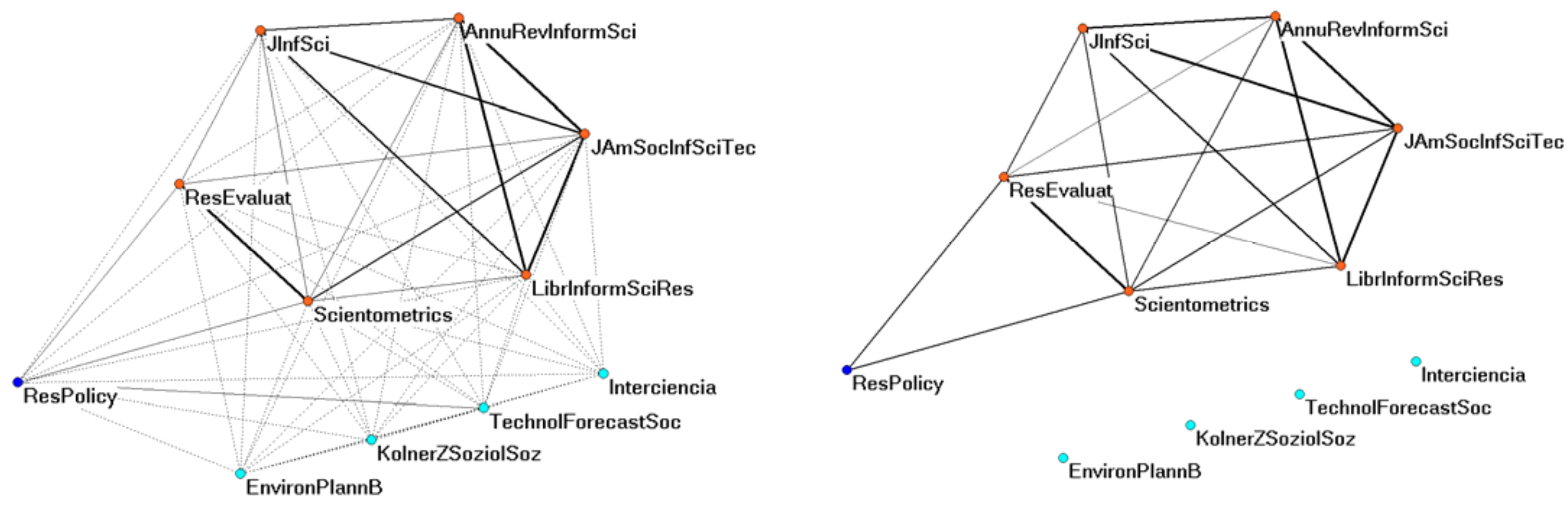

**Figure 7a and b**: Eleven journals in the citation impact environment of *Scientometrics* in 2007 with and without negative correlations in citation patterns.

Figure 7 shows the use of the upper limit of the threshold value for the cosine (according with $r = 0$) in another application. On the left side (Figure 7a), the citation impact environment ("cited patterns") of the eleven journals which cited *Scientometrics* in 2007 to the extent of more than 1% of its total number of citations in this year ($n = 1515$) is visualized using the Pearson correlation coefficients among the citation patterns. Negative values of $r$ are depicted as dashed lines.

The right-hand figure can be generated by deleting these dashed edges. However, this Figure 7b is based on using the upper limit of the cosine for $r = 0$, that is, cosine > 0.301. The use of the cosine enhances the edges between the journal *Research Policy,* on the one hand, and *Research Evaluation* and *Scientometrics,* on the other. These relations were depressed because of the zeros prevailing in the comparison with other journals in this set (Ahlgren *et al*., 2003). Thus, the use of the cosine improves on the visualizations, and the cosine value predicted by the model provides us with a useful threshold.

In summary, the use of the upper limit of the cosine which corresponds to the value of $r = 0$ can be considered conservative, but warrants focusing on the meaningful part of the network when using the cosine as similarity criterion. In the meantime, this "Egghe-Leydesdorff" threshold has been implemented in the output of the various bibliometric programs available at http://www.leydesdorff.net/software.htm for users who wish to visualize the resulting cosine-normalized matrices.

## 7. The relation between r and similarity measures other than Cos

In the introduction we noted the functional relationships between $\mathrm{Cos}$ and other similarity measures such as Jaccard, Dice, etc. Based on $L^2$-norm relations, e.g. $\left\|\vec{X}\right\|_2 = \left\|\vec{Y}\right\|_2$ (but generalizations are given in Egghe (2008)) we could prove in Egghe (2008) that ($J$ = Jaccard)

$$J = \frac{\mathrm{Cos}}{2 - \mathrm{Cos}} \tag{20}$$

and that $E = \mathrm{Cos}$ ($E$ = Dice), and the same holds for the other similarity measures discussed in Egghe (2008). It is then clear that the combination of these results with (13) yields the relations between *r* and these other measures. Under the above assumptions of $L^2$-norm equality we see, since $E = \mathrm{Cos}$, that (13) is also valid for $\mathrm{Cos}$ replaced by $E$. For $J$, using (13) and (20) one obtains:

$$\mathrm{Cos} = \frac{2J}{J+1} \tag{21}$$

and hence

$$r = \frac{n}{\sqrt{n - a^2}\sqrt{n - b^2}}\left(\frac{2J}{J+1} - \frac{ab}{n}\right) \tag{22}$$

which is a relation as depicted in Figure 8, for the first example (the asymmetric binary occurrence matrix case).

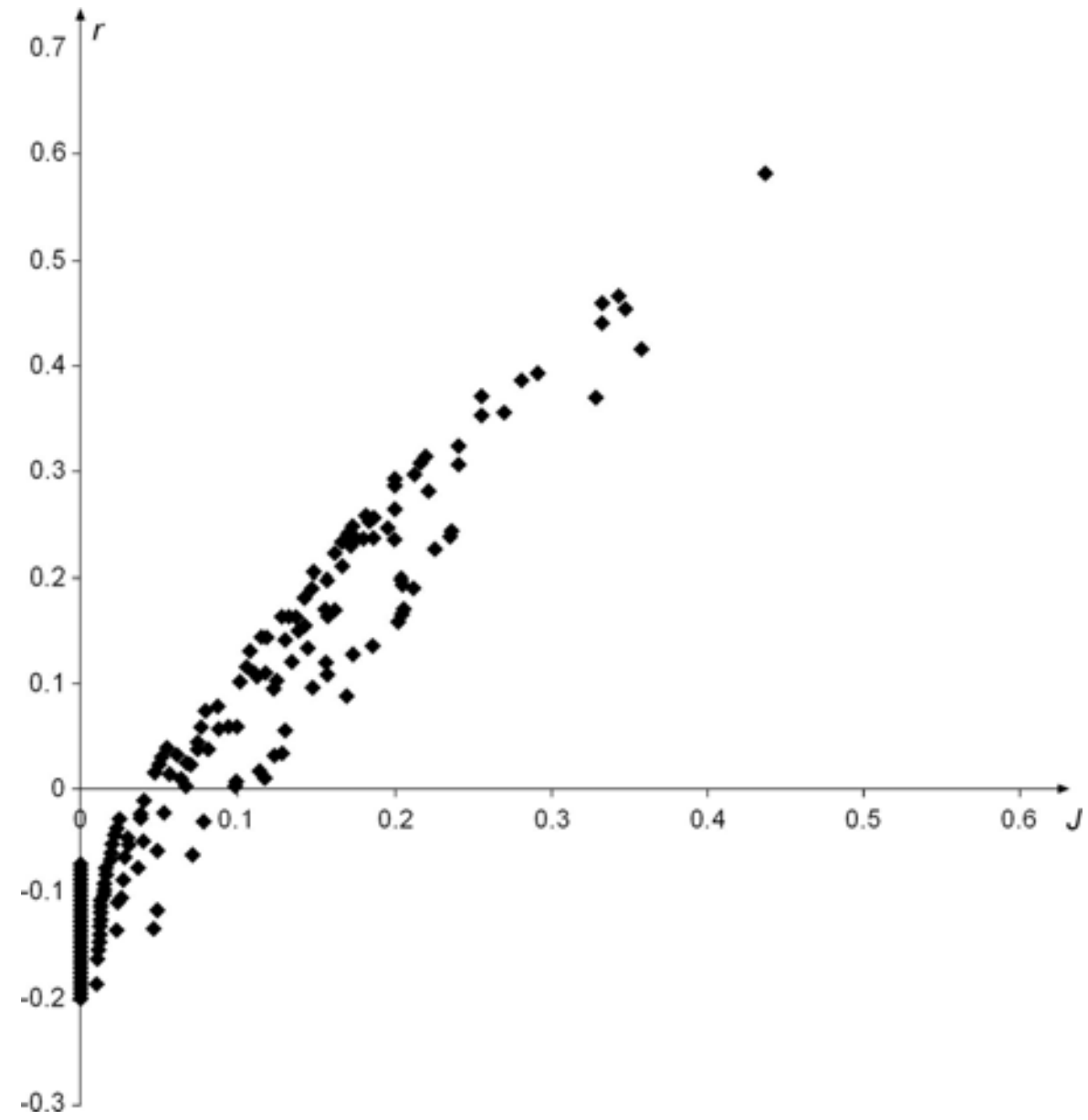

**Figure 8:** The relation between r and J for the binary asymmetric occurrence matrix

The faster increase of this cloud of points, compared with the one in Figure 2 follows from the fact that (20) implies that $J < \text{Cos}$ (since $0 \le \text{Cos} \le 1$) if $\text{Cos} \in \left]0,1\right[$: in fact $J$ is convexly increasing in $\text{Cos}$, below the first bissectrix: see Leydesdorff (2008) and Egghe (2008).

As we showed in Egghe (2008), if $\left\|\vec{X}\right\|_2 = \left\|\vec{Y}\right\|_2$ all the other similarity measures are equal to $\text{Cos}$, so that we evidently have graphs as in Figures 2 and 3 of the relation between *r* and the other measures.

## 8. Conclusion

In this paper we have presented a model for the relation between Pearson's correlation coefficient *r* and Salton's cosine measure. We have shown that this relation is not a pure function, but that the cloud of points $(\text{Cos}, r)$ can be described by a sheaf of increasing straight lines whose slopes decrease, the higher the straight line is in the sheaf. The negative part of *r* is explained, and we have explained why the *r*-range (thickness) of the cloud decreases when $\text{Cos}$ increases. All these theoretical findings are confirmed on two data sets from Ahlgren, Jarneving & Rousseau (2003) using co-citation data for 24 informetricians: vectors in the asymmetric occurrence matrix and the symmetric co-citation matrix.

The algorithm enables us to determine the threshold value for the cosine above which none of the corresponding Pearson correlation coefficients on the basis of the same data matrix will be lower than zero. In general, a cosine can never correspond with an $r < 0$, if one divides the product between the two largest values for a and b (that is, $\frac{\sum_{i=1}^{n} x_i}{\sqrt{\sum_{i=1}^{n} x_i^2}}$ for each vector) by the size of the vector n.

In the case of Table 1, for example, the two largest sumtotals in the asymmetrical matrix were 64 (for Narin) and 60 (for Schubert). Therefore, a was $\sqrt{64}$ and b was $\sqrt{60}$ and hence $\sqrt{ab}$ was $61.967734$. Since $n = 279$ in this case, the cosine should be chosen above 61.97/279 = $0.2221066$ because above this threshold one expects no single Pearson correlation to be negative. This cosine threshold value is sample (that is, n-) specific. However, one can automate the calculation of this value for any dataset by using Equation 18.